\documentclass[a4paper]{article}

\usepackage[english]{babel}
\usepackage[utf8x]{inputenc}
\usepackage[T1]{fontenc}
\usepackage{comment}
\usepackage{lscape}

\usepackage{authblk}
\usepackage{amsmath}
\usepackage{amssymb}
\usepackage{siunitx}
\usepackage[table]{xcolor}
\usepackage{graphicx}
\usepackage{orcidlink}
\usepackage{fullpage}
\usepackage[colorinlistoftodos]{todonotes}

\usepackage{color}

\makeatletter
\newcommand*\bigcdot{\mathpalette\bigcdot@{0.7}}
\newcommand*\bigcdot@[2]{\mathbin{\vcenter{\hbox{\scalebox{#2}{$\m@th#1\bullet$}}}}}
\makeatother

\title{Z-scores-based methods and their application to biological monitoring: An extended analysis of professional soccer players and cyclists athletes}
\author[1,2]{Geoffroy Berthelot~\orcidlink{0000-0003-4036-6114}}
\author[3]{Brigitte Gelein~\orcidlink{0009-0005-7201-3200}}
\author[4]{\'{E}ric Meinadier}
\author[5]{Emmanuel Orhant~\orcidlink{0000-0002-0548-3940}}
\author[6]{J\'{e}r\^{o}me Dedecker~\orcidlink{0000-0002-8838-0356}}
\affil[1]{Institut de Recherche bioM\'{e}dicale et d'Epid\'{e}miologie du Sport (IRMES, UPR7329), INSEP (Institut National du Sport, de l'Expertise et de la Performance), Paris, France}
\affil[2]{Research Laboratory for Interdisciplinary Studies (RELAIS), 89100 Sens, France}
\affil[3]{ENSAI, Institut de recherche math\'{e}matique de Rennes, UMR CNRS 6625, Rennes, France}
\affil[4]{Medical Department, French Federation of Cycling, Paris, France}
\affil[5]{French Football Federation (FFF), Clairefontaine Medical Centre, FIFA Medical Center of Excellence, Clairefontaine, France}
\affil[6]{Universit\'{e} Paris Cit\'{e}, CNRS, UMR 8145, MAP5, F-75006 Paris, France}
\date{}

\begin{document}

\maketitle

\begin{abstract}
The increase in the collection of biological data allows for the individual and longitudinal monitoring of hematological or urine biomarkers. However, identifying abnormal behavior in these biological sequences is not trivial. Moreover, the complexity of the biological data (correlation between biomarkers, seasonal effects, etc.) is also an issue. Z-score methods can help assess the abnormality in these longitudinal sequences while capturing some features of the biological complexity. This work details a statistical framework for handling biological sequences using three custom Z-score methods in the intra-individual variability scope. These methods can detect abnormal samples in the longitudinal sequences with respect to the seasonality, chronological time or correlation between biomarkers. One of these methods is an extension of one custom Z-score method to the Gaussian linear model, which allows for including additional variables in the model design. We illustrate the use of the framework on the longitudinal data of 3,936 professional soccer players (5 biomarkers) and 1,683 amateur or professional cyclists (10 biomarkers). The results show that a particular Z-score method, designed to detect a change in a series of consecutive observations, measured a high proportion of abnormal values (more than three times the false positive rate) in the ferritin and IGF1 biomarkers for both data sets. The proposed framework and methods could be applied in other contexts, such as the clinical patient follow-up in monitoring abnormal values of biological markers. The methods are flexible enough to include more complicated biological features, which can be directly incorporated into the model design.
\end{abstract}

\section{Introduction}
Blood and urinary sampling allows for the early detection of physiological or biological changes at the individual level. In personalized medicine, the longitudinal analysis of biological samples is critical for monitoring and regulating therapeutic efficacy and disease progression in cancer patients for example \cite{zenil2022immune, goetz2018personalized}. The approach has also been developed in sport where athletes' biological values are scrutinized for the detection of \textquoteleft abnormal\textquoteright (i.e. pathological or doping) conditions \cite{saugy2014monitoring, sottas2011athlete, zorzoli2014practical, sauliere2019z}. This approach was formalized in 2009 with the introduction of the Athlete Biological Passport, which is designed to indirectly detect the effects of doping \cite{zorzoli2014practical, saugy2020antidoping}. It consists of a normalized follow-up of the athletes' hematological parameters which can be analyzed to detect abnormal values. These follow-ups are essential, as Malcovati et al. \cite{malcovati2003hematologic} and Egger et al. \cite{egger2016interindividual} underlined that inter-individual variability is higher than intra-individual variability. Therefore, relying on population data to estimate intra-individual variability may introduce additional bias. Different black and white box methods have been proposed to automatically detect abnormal values in longitudinal samples of biomarkers. Black box methods, such as those used in artificial intelligence and machine learning, show promise in personalized medicine \cite{damavsevivcius2024deep} and in longitudinal biomedical data \cite{cascarano2023machine}. However, the interpretability of the resulting detections is not straightforward, due to their opaque internal mechanisms \cite{castelvecchi2016can, rudin2019stop, lipton2018mythos, li2024shapley, wallace2023use}. In addition, identifying the sources of bias or error remains a significant challenge \cite{mehrabi2021survey}. They also come at high computational costs and significant energy consumption, raising concerns about their environmental impact \cite{strubell2020energy, patterson2021carbon}. Finally, they depend on large amounts of data, which may not be easily available if one only consider individual baselines and specific biomarkers. White box approaches typically rely on the detection of an abnormal deviation from an individual\textquoteright s baseline using a statistical metric such as the Z-score \cite{curtis2016mystery}. It has found applications in biology \cite{cheadle2003analysis}, neuroimaging \cite{ishii2000diagnostic}, psychology \cite{guilford1973structure}, sport \cite{sauliere2019z} and medicine \cite{dallaire2015bias, roshan2021comparison, roshan2025adaptive} among others. The abnormal biomarker values are characterized by an explicit computation, providing a clear rationale for their deviation from the expected baseline. Additional work can then be conducted to further investigate and tune the statistics and expected bias while giving further control over the expected error types and rates. This work introduces a framework for the statistical analysis of individual and longitudinal biological data in the context of personalized medicine. It is centered on custom Z-score methods, designed to capture intra-individual variability. It also describes the results of these methods applied to two data sets containing biological data from elite athletes.

\section{Methods}
\label{Method_head}
\begin{figure}
  \centering
    \includegraphics[width=\textwidth]{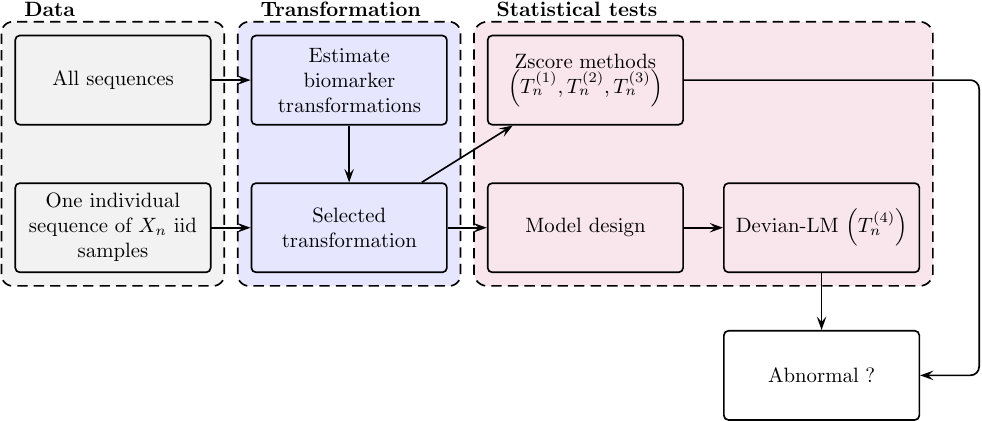}
  \caption{Block diagram of the methodological process for one biomarker only (see Section \ref{Method_head}). First, we choose from among several deterministic transformations the one for which it seems most likely that the transformed observations come from a Gaussian sample (``Transformation'' block). We apply this transformation to each individual (ie. individual sequences), before computing different Z-scores methods (``Statistical test'' block). These methods allow for measuring the abnormality of one sequence. We can then assess the proportion of abnormal sequences (individuals) in our population. The procedure is repeated for each biomarker. The correlation between different biomarkers is not pictured here, as it involves multiple biomarkers simultaneously.}
  \label{fig1:blockdiagram}
\end{figure}

The methodology section is organized as follows: we first present the two longitudinal biological datasets; we then evaluate and - if necessary - transform the data so that it is more likely to come from a Gaussian sample; we study the correlation of biomarkers;  and we apply a series of Z-score methods \cite{sauliere2019z, berthelot2025devian}. We evaluate the proportion of abnormal individuals for each Z-score method. The framework is summarized in Figure \ref{fig1:blockdiagram}.

\subsection{Soccer and cycling data}
After obtaining the approval of the Institutional Ethics Committee, two datasets of elite soccer players and elite cyclists were obtained. The soccer dataset consists of five typical biological markers from 3,936 male soccer players from the French elite leagues 1 and 2. The markers include concentrations of ferritin ($\mu$mol/L), serum iron ($\mu$mol/L), hemoglobin (g/L), erythrocytes (T/L), and hematocrit levels (\%). The biomarkers were collected every 6 months in July/August and in January/March from 2006 to 2019 for a total of 27 collections. The large interval between two measurements (around 6 months) allowed for independent samplings \cite{sharpe2006third}.

The cycling dataset consists of ten biological markers from 1,134 amateur, 389 professional cyclists and 160 with both amateur and professional status (total 1,683), all male athletes. These competed in BMX (105 athletes, all amateurs), road cycling (1,052 athletes of whom 939 amateurs and 113 professionals), cross-country mountain biking (143 athletes, all amateurs), downhill mountain biking (60 athletes, all amateurs), cyclo-cross (49 athletes, all amateurs), track sprint (38 athletes, all amateurs), bike trials (30 athletes, all amateurs), track pursuit (40 athletes, all amateurs), multiple disciplines (6 athletes, 5 amateurs and 1 professional) and both multiple disciplines and status (160 athletes). The average age of sample collection is 24.07 years-old, the youngest and oldest athletes are 13.30 and 43.54 years old, respectively. The first sample was collected on the 09th Jan 2003 and the last sample was collected on the 12th Feb 2014. Markers include concentrations of ferritin ($\mu$mol/L), hemoglobin (g/L), erythrocytes (T/L), hematocrit levels (\%), reticulocytes (G/L and \%), OFF-hr score, insulin-like growth factor 1 (IGF1) (ng/mL), cortisol (nmol/L), testosterone (nmol/L), osteocalcin (ng/mL). In this dataset, 99.92\% of the samplings are spaced at least 6 days apart.

\subsection{Normality of the sequences}
\label{Normality}
For most of the methods for detecting abnormal values that we will describe in Subsection \ref{Zsc}, we will assume that the random sample $X_1, \ldots , X_n$ (generating the observations $x_1,  \ldots , x_n$) is a sequence of iid (independent, identically distributed) random variables with  normal distribution. An assumption about the distribution of the variables seems necessary, as we are dealing with small samples, but it is nevertheless restrictive. In this context, a value (or a sequence of values) may be declared abnormal because the assumption of normality is not satisfied.
Of course, we want this scenario to occur as rarely as possible, which is why we first propose pre-processing the data, as described below. For each biomarker, we will apply a family of transformations (described at the end of this paragraph) to all  series of observations containing at least four observations (recall that, for one fixed biomarker,  there is one series of observations per individual). Then, for any transformation, we apply to all transformed sequences a Shapiro test of normality. This gives a series of $p$-values corresponding to all the individuals for which there is at least four observations in the series. If the assumption of normality was true, these $p$-values should be uniformly distributed over $[0,1]$. It is therefore natural to use a Kolmogorov-Smirnov test of adequation to the uniform distribution on these series of $p$-values: this gives a global $p$-value measuring the adequation to the uniform distribution. At this stage, for each deterministic transformation, we have a single $p$-value measuring the adequation to the uniform distribution. We then choose the transformation for which this $p$-value is maximal: this is a way to select the transformation for which the transformed individuals samples are ``globally'' closest to normally distributed samples. Of course, this procedure is not completely satisfactory, since some of the sequences of observations contains ``true'' abnormal observations, that is observations that are much too far from the other observations in the sample, so that the assumption of identical distribution is not satisfied. this explains partly why the $p$-values obtained through the Kolmogorov test are all very small (whatever the transformations used), see Table \ref{tab1:Normalite}. However, by doing so, one can limit the number of detections of abnormal values due to the fact that the variables are not distributed according to the normal distribution. To conclude, with this approach, we aim at detecting abnormal sequences from non-identically distributed random variables, rather than abnormal sequences whose underlying distribution deviates from normality.

Let us now describe the transformations that we use for our study: the identity (no transformation), $\sqrt[m]{X_i}$ (the $m$th-root transformation, with $m \in \{2,\ldots,10\}$), $\log(X_i)$ (the logarithmic transformation), the Lambert $W_0(X_i)$ function and Box-Cox transformations. For the Box-Cox transformations, we use $\lambda = \{-0.0606, 0.0202, -0.3030\}$ as reported in \cite{kang2023indirect}. These 3 values were empirically estimated on adults (hospital patients) for the C-reactive protein, the erythrocyte sedimentation rate and the presepsin respectively. Previous works reported similar transformations \cite{athanasiadou2020hyperhydration, zhang2025testosterone, medicina56080400, sezgin2020clinical} or more sophisticated ones such as the LMS method \cite{cole1992smoothing}. However these transformations are typically applied to cohorts (see \cite{stojanovic2021serum} and \cite{isojima2012standardized} for the Serbian and Japanese populations respectively) thus departing from the individual approach described in this work.

\begin{table}
\centering
{\footnotesize
\begin{tabular}{ | l | l | l | }
    \hline
    Biomarker & Soccer & Cycling\\
    \hline
    Erythrocytes (T/L)      & $\sqrt[3]{X_i}$ $\left(p = \num{2.42e-07}\right)$ & BoxCox $\gamma = -0.0606$ $\left(p = \num{1.09e-06}\right)$ \\
    Hemoglobin (g/L)        & $\sqrt[3]{X_i}$ $\left(p = \num{9.94e-13}\right)$ & $\sqrt{X_i}$ $\left(p = \num{1.95e-04}\right)$ \\
    Hematocrit (\%)         & BoxCox $\gamma = -0.0606$ $\left(p = \num{2.43e-05}\right)$ & $\sqrt{X_i}$ $\left(p = \num{1.05e-04}\right)$ \\
    Ferritin ($\mu$mol/L)   & $\sqrt[6]{X_i}$ $\left(p = \num{6.79e-16}\right)$ & $\sqrt[5]{X_i}$ $\left(p = \num{1.44e-09}\right)$ \\
    Serum iron ($\mu$mol/L) & $\sqrt[5]{X_i}$ $\left(p = \num{1.52e-09}\right)$ &  \\
    Reticulocyte (G/L) & & $W_0(X_i)$ $\left(p = \num{6.09e-06}\right)$ \\
	Reticulocyte (\%)  & & BoxCox $\gamma = -0.0202$ $\left(p = \num{3.32e-10}\right)$\\
	OFF-hr & & $X_i^2$ $\left(p = \num{1.68e-02}\right)$ \\
	IGF1 (ng/mL) & & BoxCox $\gamma = -0.0303$ $\left(p = \num{5.01e-09}\right)$\\
	Cortisol (nmol/L) & & $\sqrt{X_i}$ $\left(p = \num{2.59e-08}\right)$ \\
	Testosterone (nmol/L) & & $\sqrt{X_i}$ $\left(p = \num{1.03e-05}\right)$ \\
	Osteocalcin (ng/mL) & & $\sqrt[8]{X_i}$ $\left(p = \num{8.41e-01}\right)$\\
   \hline
\end{tabular}
\caption{\label{tab1:Normalite} The selected transformations detailed for each of the two datasets. The $p$-values of the Kolmogorov-Smirnov tests are given for each selected transformation (see Subsection \ref{Normality}).}
}
\end{table}

\subsection{Correlation between biomarkers}
\label{Correlation}
\begin{figure}
  \centering
    \includegraphics[width=\textwidth]{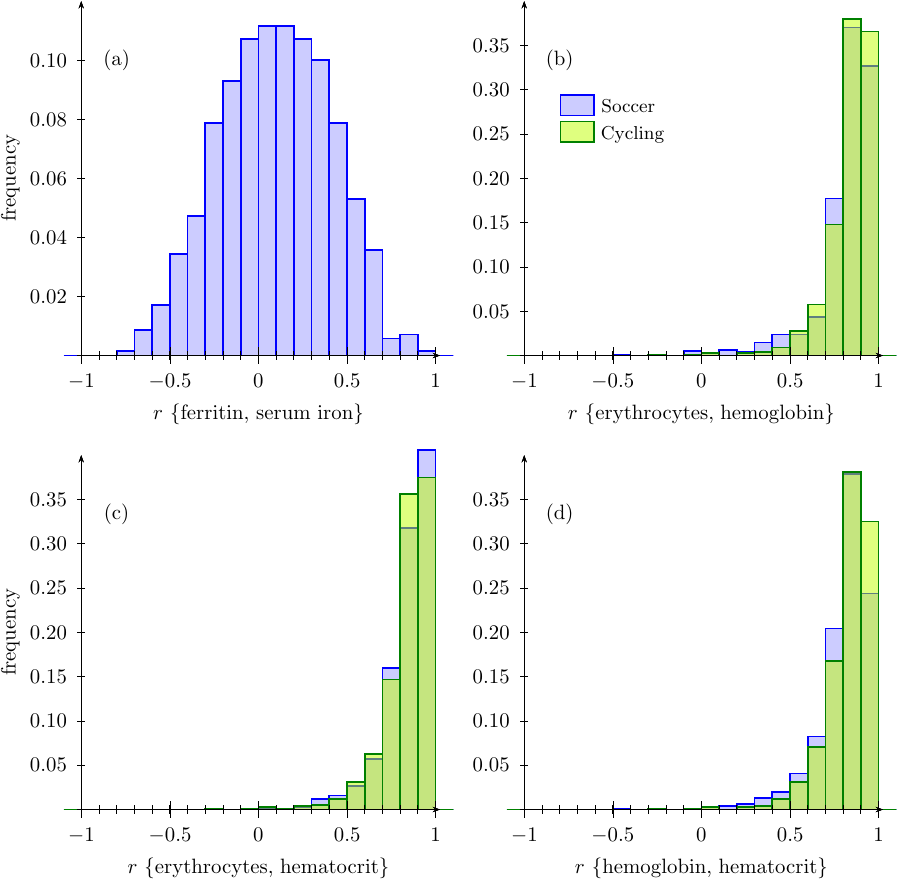}
    \caption{Histograms of the empirical Pearson correlation coefficients $r$ between biomarkers. In \textbf{(a)} the ferritin and serum iron biomarkers are only available for the soccer dataset.}
    \label{fig2:correlationBioMakers}
\end{figure}

We measure the correlation coefficients $r$ for each sequence and for the following couples of biomarkers: \{ferritin, serum iron\} (soccer only); \{erythrocytes, hemoglobin\}; \{erythrocytes,  hematocrit\}; \{hemoglobin,  hematocrit\}. For this analysis, we only include individuals for whom there are at least 10 observations ($n \ge 10$). The distribution of the correlation coefficients $r$ are presented in Figure \ref{fig2:correlationBioMakers}.

\subsection{Z-score background} \label{Zsc}
Given a new biomarker observation $x_n$, how to determine if it is abnormal?  A typical method is to compare the new value $x_n$ to the average of the other (past) values in a sequence of observations $x_1, x_2, \ldots , x_n$ from the same individual. One can then define the statistic:
\begin{equation}\label{Zscore_0}
    T_n^{(0)} = \dfrac{X_n - \bar{X}_{n -1}}{ \hat{\sigma}_{n-1} \sqrt{1+\dfrac{1}{n-1}} }
\end{equation}
where $\bar{X}_{n-1}$ and $\hat{\sigma}_{n-1}$ are the empirical mean and variance of the random sequence $X_1,\ldots,X_{n-1}$ (ie. the past  $n-1$ random variables):
\begin{equation}\label{musigma}
    \bar{X}_{n -1} = \dfrac{1}{n-1} \sum_{k=1}^{n-1} X_k \qquad \text{and} \qquad \hat{\sigma}_{n-1} = \dfrac{1}{n-2} \sum_{k=1}^{n-1} \left(X_k - \bar{X}_{n -1}\right)^2
\end{equation}
Assuming that the variables $(X_i)_{1\leq i \leq n-1}$ are iid  with common $\mathcal{N}(\mu,\,\sigma^{2})$ distribution, then under $H_0$ ``$X_n$ has the same $\mathcal{N}(\mu,\,\sigma^{2})$ distribution'', the test statistic $T_n^{(0)}$ has the Student$(n-2)$ distribution. In order to test (at level $\alpha$) if the new observation $x_n$ is abnormal, the procedure is  then to compare the observed value  $|t_n^{(0)}|$ to  the quantile $t_{\alpha/2, n-2}$ of order $1-\alpha/2$ of the Student($n$ - 2) distribution. More precisely, we reject $H_0$ if $|t_n^{(0)}| > t_{\alpha/2, n-2}$.

This is the simplest of the $Z$-score type procedures, for which the test statistic distribution is comppletely known under $H_0$. In \cite{sauliere2019z}, we developed other, more general $Z$-score type statistics in order to answer a set of more complex questions. For these other procedures, the distributions of the test statistics are free (i.e. do not depend on the  unknown parameters $\mu$ and $\sigma^2$) but they have no explicit expression and must therefore be tabulated using a Monte Carlo method. We will briefly describe these methods in the following paragraphs.

\subsubsection{Z-score method for detecting an abnormal observation in a sequence}
In order to check if there is an abnormal observation $x_i$ (not necessarily the last one) in the sequence $x_1, x_2, ..., x_n$, Sauli\`{e}re et al.  \cite{sauliere2019z} introduced the following extension of the statistic $T_n^{(0)}$ (see \eqref{Zscore_0}):
\begin{equation}\label{Zscore_1}
T_n^{(1)} = \max_{i \in \{1, \ldots , n\}} \left | \frac {X_{i}-\bar X_{n, -i}}
{\hat \sigma_{n, -i}\sqrt{1 +\frac{1}{n-1}}} \right | \, ,
\end{equation}
where
\begin{equation}\label{musigma2}
\bar X_{n, -i} = \frac{1}{n-1} \sum_{k=1, k\neq i}^n X_k
\quad \text{and} \quad
\hat \sigma_{n, -i}^2 = \frac{1}{n-2} \sum_{k=1, k \neq i}^n (X_k - \bar X_{n, -i})^2 \, .
\end{equation}
with $\bar X_{n, -i}$ and $\hat \sigma^2_{n-1}$ the empirical mean and variance of the sample $X_1, \ldots, X_{n}$ without the $i$th observation (denoted $-i$). This method is based on the maximum of $n$ (non independent) variables with Student($n-2$) distribution.

\subsubsection{Z-score method for detecting an abnormal subsequence}
An extension of the statistic $T_n^{(1)}$ (see \eqref{Zscore_1}) was introduced to detect a series of consecutive observations that deviate from the rest of the series (see Section 2.3, \textquoteleft Method 2 \textquoteright~in \cite{sauliere2019z}):
\begin{equation}\label{Zscore_2}
T_n^{(2)} = \max_{I \in {\mathcal I}} \left | \frac {\bar X_{I}-\bar X_{\bar I}}
{\hat \sigma_{n, I}\sqrt{\frac{1}{|I|} +\frac{1}{n-|I|}}} \right |
\end{equation}
where
${\mathcal I}$ is the collection of all possible intervals $I$ included in $\{1, \ldots, n\}$ with length $1\leq |I|<n$,
\begin{equation}\label{mum2}
\bar X_{I} = \dfrac{1}{|I|} \sum_{k \in I} X_k\, , \quad
\bar X_{\bar I} = \dfrac{1}{n-|I|} \sum_{k \notin I} X_k \, ,
\end{equation}
where $\bar X_{I}$ is the empirical mean on the interval $I$, $\bar X_{\bar I}$ the empirical mean on the set $\bar I$ (the complementary set of $I$) and
\begin{equation}\label{sigmam2}
\hat \sigma_{n,I}^2 = \frac{1}{n-2} \left( \sum_{k\in I} (X_k - \bar X_{I})^2 +\sum_{k\notin I}^n (X_k - \bar X_{\bar I})^2 \right) \, .
\end{equation}
is the empirical variance with respect to the interval $I$.

\subsubsection{Multivariate extension}
A multivariate extension of the statistic $T_n^{(1)}$ was also introduced (see Subsection 2.2, \textquoteleft Method 1\textquoteright~ multivariate extension in \cite{sauliere2019z}). Our problem is now  to test the abnormality of multiple and correlated biomarkers at once. To deal with this situation, we assume that the observations are  obtained from  $n$ independent Gaussian ${\mathbb R}^d$-valued random variables $\boldsymbol{X}_1, \ldots, \boldsymbol{X}_n$  (whose correlation is described by a correlation matrix $C$, assumed invertible). In this context, we consider the natural extension of the square of the statistic $T_n^{(1)}$:
\begin{equation}\label{Zscore_1multi}
T_n^{(3)} = \frac{(n-1)}{nd}\max_{i \in \{1, \ldots , n\}}   ({\boldsymbol{X}_{i}-\bar{\boldsymbol{X}}_{n, -i}})' C_{n,-i}^{-1} ({\boldsymbol{X}_{i}-\bar{\boldsymbol{X}}_{n, -i}}) \, ,
\end{equation}
where
\begin{equation}\label{musigma1multi}
\bar{\boldsymbol{X}}_{n, -i}= \frac{1}{n-1} \sum_{k=1, k\neq i}^n \boldsymbol{X}_k
\quad \text{and} \quad
 C_{n,-i}= \frac{1}{n-1-d} \sum_{k=1, k \neq i}^n (\boldsymbol{X}_k - \bar{\boldsymbol{X}}_{n, -i}) (\boldsymbol{X}_k - \bar{\boldsymbol{X}}_{n,-i})' \, .
\end{equation}

According to the results of Subsection \ref{Correlation}, we will compute the statistic $T_n^{(3)}$  for the following (transformed) biomarker tuples: \{ferritin, serum iron\} (soccer only); \{erythrocytes, hemoglobin, hematocrit\}; (for soccer and cycling). The selected transformations for each biomarkers are identical to those used previously (see Section \ref{Normality} and Table \ref{tab1:Normalite}).

\subsubsection{DevianLM and model designs}
\label{DevianLM}
The DevianLM package implements an extension of $T_n^{(1)}$ to the Gaussian Linear Model \cite{berthelot2025devian}, where we assume that the observations are obtained from the random variables $X_1,\ldots, X_n$ such that
\begin{equation}\label{LM}
  X_i= (M \theta)_i + \varepsilon_i \, ,
\end{equation}
where $M$ is the $n\times p$ observed design matrix (assumed invertible), $\theta$ is a ${\mathbb R}^p$  vector of unknown parameters, and $\varepsilon_1, \ldots, \varepsilon_n$ is a sequence of iid random variables with common distribution ${\mathcal N}(0, \sigma^2)$, independent of the design matrix $M$.

To detect the abnormal values from the (external) studentized residuals of the model $\hat e_i(X)$, we use the statistic:
\begin{equation}\label{DevianLM}
    T_n^{(4)} = \max_{i \in \{1, \ldots , n\}} \left | \frac {X_i-\hat X_{n,i}}
    {\hat \sigma_{n, i}\sqrt{1 +L_i (M'_{(i)} M_{(i)})^{-1}L_i'}} \right |= \max_{i \in \{1, \ldots , n\}} \left |\hat e_i(X) \right | \, .
\end{equation}
where $M(i)$ is the $(n-1) \times p$ matrix $M$ deprived of its $i$th row, $L_i$ is the $i$th row of the matrix $M$, $\hat \theta_{n,i}$ is the least square estimator of $\theta$ based on the variables $(X_j, L_j)$ for $j \neq  i$,   $\hat X_{n,i}$ is the prediction of $X_i$ obtained from  $\hat \theta_{n,i}$, that is  $\hat X_{n,i} = L_i \hat \theta_{n,i}$, and $\hat \sigma_{n, i}$ is the estimator of the variance based on the variables based on the variables $(X_j, L_j)$ for $j \neq i$.

It is proved in \cite{berthelot2025devian} that the distribution of $T_n^{(4)}$ does not depend on the unknown parameters $(\theta, \sigma^2)$, but it depends on the observed matrix $M$. Hence it has to be tabulated via Monte Carlo simulations.

In our practical situation,  for the soccer and cyclists datasets, we will consider three different designs, labeled A, B, C.
The first model (model A) describes a fluctuation around a fixed baseline $\beta_0$, meaning that the equation \eqref{LM} is simply
\begin{equation}\label{ModelA}
    X_i = \beta_0 +  \varepsilon_i
\end{equation}
For this model, the statistics $T_n^{(4)}$ is exactly equal to $T_n^{(1)}$ (see \eqref{Zscore_1}).

The second model (model B) describes a fluctuation around two distinct baselines, which is equivalent to an Homoscedastic ANOVA model with two groups. For our datasets, the two groups will consist of biomarker observations in summer or winter (each season covering a period of six months). Note that the soccer dataset followed an explicit summer/winter sampling procedure while the cycling data sets recorded the exact sampling dates. For the cycling dataset we then separate the dates in two seasons: summer, from the 03/20 to the 09/22, and the remaining dates are classified as the winter season. For model B, the equation \eqref{LM} can be written as follows:
\begin{equation}\label{ModelB}
    X_i = \beta_0 + \beta_1 s_i+  \varepsilon_i
\end{equation}
where $s_i$ is a binary variable that denotes the season ($1$ for summer and $0$ for winter).

The third model (model C) is associated with the exact dates of sampling. It can only be applied to the cycling database, as the soccer database does not contain the exact date of collection. For Model C, the equation \eqref{LM} writes:
\begin{equation}\label{ModelC}
    X_i = \beta_0 + \beta_1 t_i+  \varepsilon_i
\end{equation}
where $t_i$ is the exact date of sampling of the $i$th observation of the biomarker.

\subsubsection{Application to the dataset}
We apply the statistics $T_n^{(2)}$, $T_n^{(3)}$ and $T_n^{(4)}$ (models A, B, C) to each individual sequence in the two datasets. As mentioned above, the statistics $T_n^{(4)}$ for Model A is exactly equal to $T_n^{(1)}$. The sample size $n$ of each sequence of observations should meet specific conditions. For DevianLM, this condition directly depends on the model matrix $M$ (a necessary condition being that $n > p + 1$, where $p$ is the number of columns of $M$). More precisely, these conditions are:
\begin{itemize}
  \item $T_n^{(2)}$ can be applied for sequences with $n \ge 4$ observations,
  \item $T_n^{(3)}$ (multivariate case) can be applied for sequences with $n \ge d + 2$ observations, where $d$ is the number of different biomarkers,
  \item $T_n^{(4)}$ (Model A) can be applied for sequences with $n \ge 3$ observations,
  \item $T_n^{(4)}$ (Model B) can be applied for sequences with $n_1, n_2 \ge 2$, where $n_1$ is the number of observations for the summer period, and $n_2$ is the number of observations for the winter period,
  \item $T_n^{(4)}$ (Model C) can be applied for $n \ge 4$ observations.
\end{itemize}

\section{Results}
The proportion of detected abnormal sequences (or equivalently the proportion of abnormal individuals, since there is one sequence of observation per individual) is presented in Fig. \ref{fig3:detection1} and Fig. \ref{fig4:detection2} for statistics $T_n^{(2)}$, $T_n^{(3)}$ and $T_n^{(4)}$ (Model A, B, C) respectively, for the two datasets. The proportion of abnormal sequences exceed the expected false-positive rate in most of the cases, especially for the statistic $T_n^{(2)}$ with values as high as 26.67\% (cycling) and 18.67\% (soccer) for the ferritin, and 35.97\% (cycling) for the IGF1. We also detail the results by athletes' characteristics for the cycling dataset below.

\subsection{Abnormal sequences per status (cycling)}
The proportion of abnormal sequences for each statistic, with respect to the athletes' status (amateur or professional), is given by biomarker in Tab. \ref{tab2:propAmaPro} for all the statistics except $T_n^{(3)}$ (the multivariate extension). For this statistic, the proportion of abnormal sequences for the erythrocytes \& hemoglobin \& hematocrit biomarkers is 53/604 (8.77\%) (amateurs) and 27/160 (16.88\%) (professionals). The proportion for athletes with multiple status (ie. athletes who switched from the amateur status to the professional one) are given below:
\begin{itemize}
    \item \textbf{Erythrocytes}: $T_n^{(2)}$ 24/158 (15.19\%) ; $T_n^{(4)}$ (A) 4/160 (2.50\%) ; $T_n^{(4)}$ (B) 11/158 (6.96\%); $T_n^{(4)}$ (C) 9/158 (5.70\%)
    \item \textbf{Hemoglobin}: $T_n^{(2)}$ 21/158 (13.29\%) ; $T_n^{(4)}$ (A) 9/160 (5.63\%); $T_n^{(4)}$ (B) 11/158 (6.96\%); $T_n^{(4)}$ (C) 11/158 (6.96\%)
    \item \textbf{Hematocrit}: $T_n^{(2)}$ 16/158 (10.13\%) ; $T_n^{(4)}$ (A) 11/160 (6.88\%); $T_n^{(4)}$ (B) 13/158 (8.23\%); $T_n^{(4)}$ (C) 14/158 (8.86\%)
    \item \textbf{Ferritin}: $T_n^{(2)}$ 89/158 (56.33\%) ; $T_n^{(4)}$ (A) 7/160 (4.38\%); $T_n^{(4)}$ (B) 20/158 (12.66\%); $T_n^{(4)}$ (C) 22/158 (13.92\%)
    \item \textbf{Reticulocytes}: $T_n^{(2)}$ 25/154 (16.23\%) ; $T_n^{(4)}$ (A) 12/158 (7.59\%); $T_n^{(4)}$ (B) 18/155 (11.61\%); $T_n^{(4)}$ (C) 17/155 (10.97\%)
    \item \textbf{Reticulocytes (\%)}: $T_n^{(2)}$ 25/154 (16.23\%) ; $T_n^{(4)}$ (A) 11/158 (6.96\%); $T_n^{(4)}$ (B) 15/155 (9.68\%); $T_n^{(4)}$ (C) 15/155 (9.68\%)
    \item \textbf{OFF-hr}: $T_n^{(2)}$ 24/154 (15.58\%) ; $T_n^{(4)}$ (A) 15/158 (9.49\%); $T_n^{(4)}$ (B) 21/155 (13.55\%); $T_n^{(4)}$ (C) 18/155 (11.61\%)
    \item \textbf{IGF1}: $T_n^{(2)}$ 96/135 (71.11\%) ; $T_n^{(4)}$ (A) 11/156 (7.05\%); $T_n^{(4)}$ (B) 11/150 (7.33\%); $T_n^{(4)}$ (C) 22/150 (14.67\%)
    \item \textbf{Cortisol}: $T_n^{(2)}$ 30/158 (18.99\%) ; $T_n^{(4)}$ (A)  14/160 (8.75\%); $T_n^{(4)}$ (B) 22/158 (13.92\%); $T_n^{(4)}$ (C) 17/158 (10.76\%)
    \item \textbf{Testosterone}: $T_n^{(2)}$ 28/156 (17.95\%) ; $T_n^{(4)}$ (A) 12/160 (7.50\%); $T_n^{(4)}$ (B) 14/159 (8.81\%); $T_n^{(4)}$ (C) 12/159 (7.55\%)
    \item \textbf{Osteocalcin}: $T_n^{(2)}$ 6/53 (11.32\%) ; $T_n^{(4)}$ (A) 6/72 (8.33\%); $T_n^{(4)}$ (B) 3/68 (4.41\%); $T_n^{(4)}$ (C) 6/68 (8.82\%)
\end{itemize}

For athletes with multiple status, the statistic $T_n^{(3)}$ for the erythrocytes \& hemoglobin \& hematocrit biomarkers is 26/156 (16.67\%).

\subsection{Abnormal sequences per discipline (cycling)}
Likewise, the proportion of abnormal sequences with respect to the athletes' discipline is given in the following text, for each statistics. The number of abnormal sequences per discipline for the statistics $T_n^{(2)}$, $T_n^{(4)}$ (model A), $T_n^{(4)}$ (model B) and $T_n^{(4)}$ (model C) are given in Fig. \ref{fig5:detectedDisciplines}. The number of abnormal sequences per discipline for the statistic $T_n^{(3)}$ is:

\begin{itemize}
    \item \textbf{Erythrocytes \& Hemoglobin \& Hematocrit}: \textit{BMX} 1/39 (2.56\%) ; \textit{cyclo-cross} 0/17 (0.00\%) ; \textit{track pursuit} 4/29 (13.79\%) ; \textit{road cycling} 57/550 (10.36\%) ; \textit{downhill mountain biking} 6/29 (20.69\%) ; \textit{mountain bike trials} 2/14 (14.29\%) ; \textit{cross-country mountain biking} 7/63 (11.11\%) ; \textit{track sprint} 3/23 (13.04\%)
\end{itemize}

And for the athletes with multiple disciplines, the proportions are given for each statistic:

\begin{itemize}
    \item \textbf{Erythrocytes}: $T_n^{(2)}$ 2/14 (14.29\%) ; $T_n^{(4)}$ (A) 0/14 (0.00\%) ; $T_n^{(4)}$ (B) 0/14 (0.00\%) ; $T_n^{(4)}$ (C) 0/14 (0.00\%)
    \item \textbf{Hemoglobin}: $T_n^{(2)}$ 1/14 (7.14\%) ; $T_n^{(4)}$ (A) 1/14 (7.14\%) ; $T_n^{(4)}$ (B) 0/14 (0.00\%) ; $T_n^{(4)}$ (C) 1/14 (7.14\%)
    \item \textbf{Hematocrit}: $T_n^{(2)}$ 2/14 (14.29\%) ; $T_n^{(4)}$ (A) 1/14 (7.14\%) ; $T_n^{(4)}$ (B) 1/14 (7.14\%) ; $T_n^{(4)}$ (C) 2/14 (14.29\%)
    \item \textbf{Ferritin}: $T_n^{(2)}$ 9/14 (64.29\%) ; $T_n^{(4)}$ (A) 1/14 (7.14\%) ; $T_n^{(4)}$ (B) 0/14 (0.00\%) ; $T_n^{(4)}$ (C) 2/14 (14.29\%)
    \item \textbf{Reticulocyte}: $T_n^{(2)}$ 1/14 (7.14\%) ; $T_n^{(4)}$ (A) 0/14 (0.00\%) ; $T_n^{(4)}$ (B) 1/14 (7.14\%) ; $T_n^{(4)}$ (C) 17/155 0/14 (0.00\%)
    \item \textbf{Reticulocyte (\%)}: $T_n^{(2)}$ 1/14 (7.14\%) ; $T_n^{(4)}$ (A) 1/14 (7.14\%) ; $T_n^{(4)}$ (B) 2/14 (14.29\%) ; $T_n^{(4)}$ (C) 1/14 (7.14\%)
    \item \textbf{OFF-hr}: $T_n^{(2)}$ 4/14 (28.57\%) ; $T_n^{(4)}$ (A) 0/14 (0.00\%) ; $T_n^{(4)}$ (B) 2/14 (14.29\%) ; $T_n^{(4)}$ (C) 2/14 (14.29\%)
    \item \textbf{IGF1}: $T_n^{(2)}$ 10/12 (83.33\%) ; $T_n^{(4)}$ (A) 0/13 (0.00\%) ; $T_n^{(4)}$ (B) 1/13 (7.69\%) ; $T_n^{(4)}$ (C) 2/13 (15.38\%)
    \item \textbf{Cortisol}: $T_n^{(2)}$ 6/14 (42.86\%); $T_n^{(4)}$ (A) 0/14 (0.00\%) ; $T_n^{(4)}$ (B) 2/14 (14.29\%) ; $T_n^{(4)}$ (C) 0/14 (0.00\%)
    \item \textbf{Testosterone}: $T_n^{(2)}$ 3/13 (23.08\%) ; $T_n^{(4)}$ (A) 1/13 (7.69\%) ; $T_n^{(4)}$ (B) 2/13 (15.38\%) ; $T_n^{(4)}$ (C) 2/13 (15.38\%)
    \item \textbf{Osteocalcin}: $T_n^{(2)}$ 1/4 (25.00\%) ; $T_n^{(4)}$ (A) 1/8 (12.50\%) ; $T_n^{(4)}$ (B) 1/8 (12.50\%) ; $T_n^{(4)}$ (C) 1/8 (12.50\%)
\end{itemize}

\begin{landscape}
\begin{table}
\centering
\begin{tabular}{ | l | l | l | l | l | | l | l | l | l | }
    \hline
     & \multicolumn{4}{c||}{Amateur} & \multicolumn{4}{c|}{Professional}\\
    \hline
    Biomarker               & $T_n^{(2)}$ & $T_n^{(4)}$ (A) & $T_n^{(4)}$ (B) & $T_n^{(4)}$ (C) &
                              $T_n^{(2)}$ & $T_n^{(4)}$ (A) & $T_n^{(4)}$ (B) & $T_n^{(4)}$ (C)\\
    \hline
    Erythrocytes     & 87/681 (12.78) & 55/917 (6.00) & 23/681 (3.38) & 36/681 (5.29) & 6/72 (8.33)  & 4/81 (4.94)  & 4/72 (5.56) & 7/72 (9.72)\\
    Hemoglobin       & 65/681 (9.54)  & 76/917 (8.29) & 26/681 (3.82) & 33/681 (4.85) & 3/73 (4.11)  & 5/81 (6.17)  & 4/73 (5.48) & 3/73 (4.11) \\
    Hematocrit       & 63/680 (9.26)  & 50/917 (5.45) & 24/681 (3.52) & 28/681 (4.11) & 5/73 (6.85)  & 6/81 (7.41)  & 2/73 (2.74) & 5/73 (6.85) \\
    Ferritin         & 124/678 (18.29)& 46/915 (5.03) & 40/678 (5.90) & 56/678 (8.26) & 30/75 (40.00)& 5/83 (6.02)  & 6/75 (8.00) & 9/75 (12.00)\\
    Reticulocyte     & 69/656 (10.52) & 62/894 (6.94) & 29/656 (4.42) & 47/656 (7.16) & 8/73 (10.96) & 7/81 (8.64)  & 6/72 (8.33) & 5/72 (6.94)\\
	Reticulocyte (\%)& 72/656 (10.98) & 96/894 (10.74)& 33/656 (5.03) & 37/656 (5.64) & 9/73 (12.33) & 10/81 (12.35)& 4/72 (5.56) & 5/72 (6.94) \\
	OFF-hr           & 76/656 (11.59) & 64/894 (7.16) & 36/656 (5.49) & 51/656 (7.77) & 16/73 (21.92)& 7/81 (8.64)  & 18/72 (25.00)& 17/72 (23.61)\\
	IGF1             & 123/486 (25.31)& 31/646 (4.80) & 18/473 (3.81) & 35/473 (7.40) & 31/74 (41.89)& 2/81 (2.47)  & 4/72 (5.56) & 2/72 (2.78) \\
	Cortisol         & 75/676 (11.09) & 69/913 (7.56) & 38/676 (5.62) & 48/676 (7.10) & 17/74 (22.97)& 6/84 (7.14)  & 12/74 (16.22)& 12/74 (16.22)\\
	Testosterone     & 49/485 (10.10) & 48/666 (7.21) & 23/485 (4.74) & 33/485 (6.80) & 9/78 (11.54) & 4/84 (4.76)  & 3/75 (4.00) & 6/75 (8.00)\\
	Osteocalcin      & 11/173 (6.36)  & 15/226 (6.64) & 6/163 (3.68)  & 11/163 (6.75) & 5/61 (8.20)  & 2/66 (3.03)  & 2/56 (3.57) & 0/56 (0.00)\\
    \hline
\end{tabular}
\caption{\label{tab2:propAmaPro} Proportion of abnormal sequences for each statistic with respect to the athletes' status. For each cell of the table, the format of the numbers reads: \#positive sequences / \#total sequences (value in \%). Additionally, the proportions for $T_n^{(3)}$ and for the mixed status (amateur / professional) are given in the results section.}
\end{table}
\end{landscape}

\begin{figure}
  \centering
    \includegraphics[width=\textwidth]{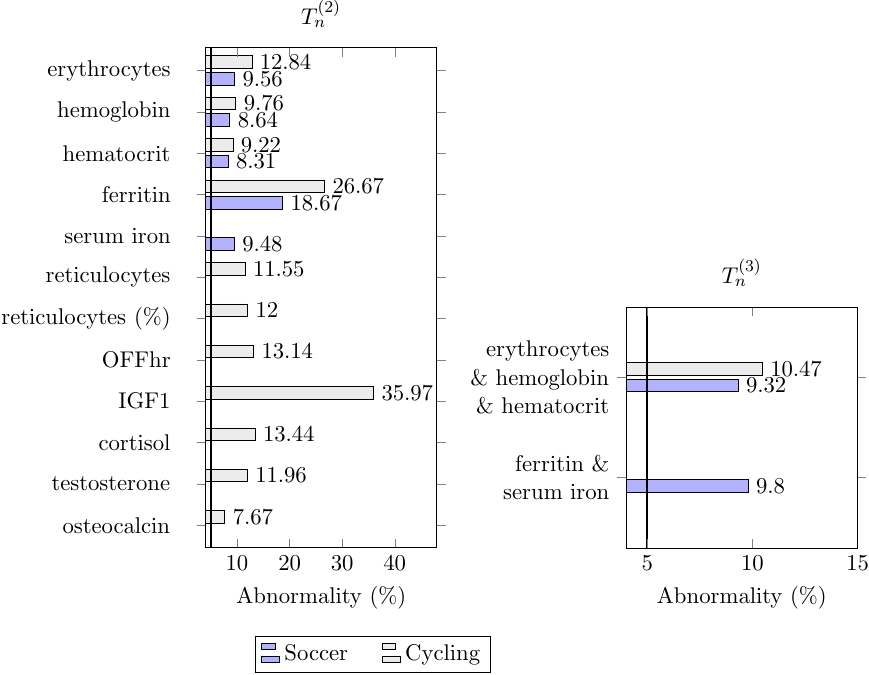}
    \caption{Detected abnormal sequences (individuals) for statistics $T_n^{(2)}$ and $T_n^{(3)}$. Each bar shows the proportion of detected abnormal sequences for the corresponding biomarker or biomarker tuple. The vertical black line indicates the 5\% significance level ($\alpha=0.05$), which corresponds to the chosen and expected false-positive rate.}
  \label{fig3:detection1}
\end{figure}

\begin{figure}
  \centering
    \includegraphics[width=\textwidth]{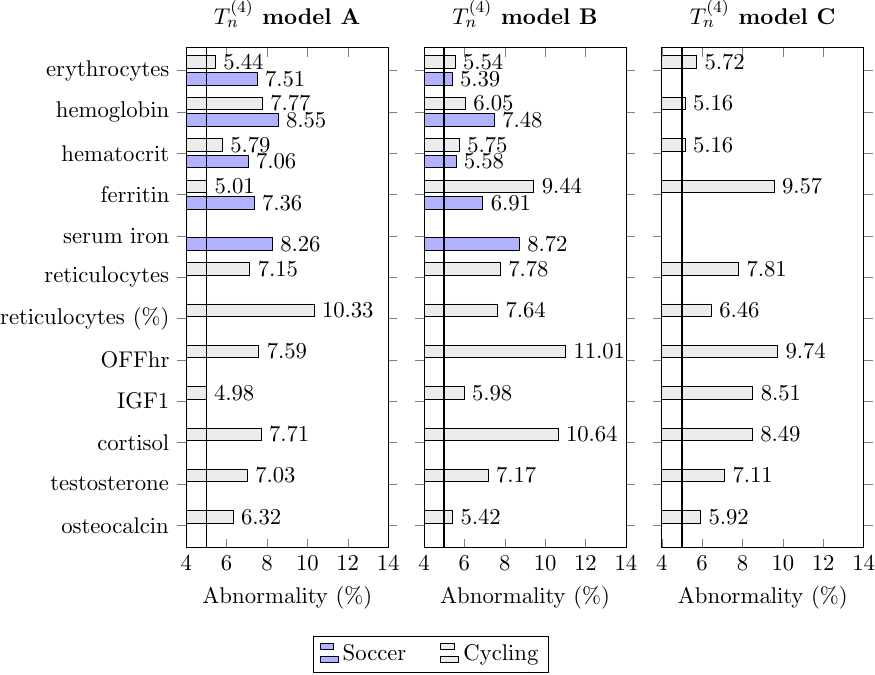}
    \caption{Detected abnormal sequences (individuals) for statistics $T_n^{(4)}$ (Model A, B, C). Each bar shows the proportion of detected abnormal sequences for the corresponding biomarker. The vertical black line indicates the 5\% significance level ($\alpha=0.05$), which corresponds to the chosen and expected false-positive rate.}
  \label{fig4:detection2}
\end{figure}

\begin{figure}
  \centering
    \includegraphics[width=\textwidth]{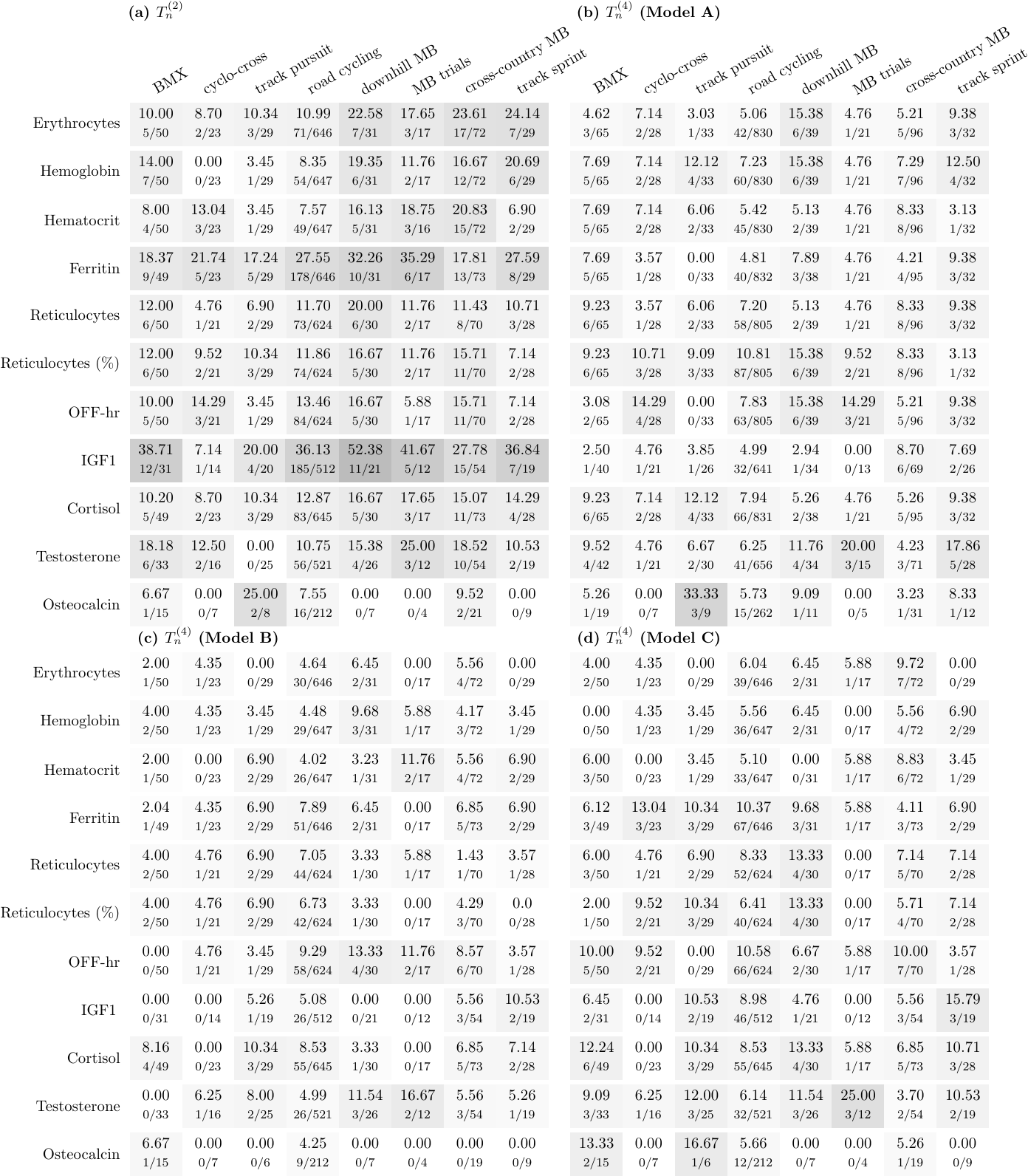}
    \caption{Heat map of the detected abnormal sequences (individuals) for statistics $T_n^{(2)}$ (panel \textbf{(a)}) and $T_n^{(4)}$ (Model A, B, C, panel \textbf{(b)-(d)}) per discipline. The number in each cell is the percentage, and the proportion of detected individuals is given below each percentage as \#detected sequences / \#sequences. The letters 'MB' stand for 'Mountain Biking'.}
  \label{fig5:detectedDisciplines}
\end{figure}

\begin{figure}
  \centering
    \includegraphics[width=\textwidth]{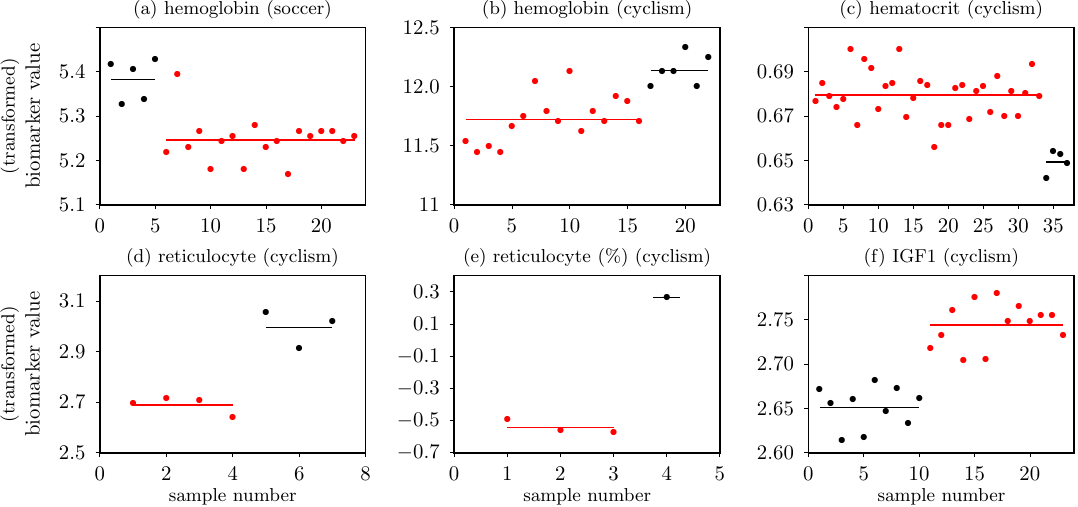}
    \caption{Example of detection for the statistic $T_n^{(2)}$. This figure illustrates the detection of a change in 6 sequences, randomly chosen. The detected change in the sequence are indicated by black dots vs. red dots. In all the 6 panels, the data showed is the transformed data (see section \ref{Normality} and Fig. \ref{fig1:blockdiagram}). The colored lines indicate the average value of each subsequence.}
  \label{fig5:exampleMethod2}
\end{figure}

The proportions of abnormal values for the statistic $T_n^{(2)}$ for the ferritin and IGF1 biomarkers are particularly high. It is observable for all disciplines but occurs more specifically among professionals (see Table \ref{tab2:propAmaPro}). Some examples of abnormal sequences detected by this statistic are provided in Fig. \ref{fig5:exampleMethod2}.

\section{Discussion}
In this article, we implement three different methods for detecting abnormal values in real data sets composed of professional or amateur athletes (cyclists or soccer players), on whom several samples of different biomarkers were taken. These methods detect different abnormalities, ranging from detecting a single outlier ($T_n^{(4)}$ model A) to a subsequence of outliers ($T_n^{(2)}$), or integrating the correlation between biomarkers ($T_n^{(3)}$) or the time component ($T_n^{(4)}$ model B and C). All these methods allow for taking into account different characteristics of the longitudinal intra-individual data. In some detections, the proportion of abnormal sequences far exceeds the expected significance level (Fig. \ref{fig3:detection1} and \ref{fig4:detection2}). In particular, the statistic $T_n^{(2)}$ detects more abnormal sequences compared to the other methods. This method differs from the other methods presented here, and can also be classified as a ``change point detection'' method. It therefore allows outliers to be detected, as well as groups of observations that fluctuate around a reference value that differs from that of the other observations (see Fig. \ref{fig5:exampleMethod2}). This allows for greater sensitivity compared to the other methods. Our study suggests that this method should be preferred to other methods for detecting changes in longitudinal follow-ups.

The large number of abnormal ferritin sequences detected can be the result of temporary biological changes due to repeated endurance training \cite{clenin2015iron}, inflammation/infection \& iron loss and hemolysis \cite{namaste2017adjusting, peeling2010exercise, keller2024iron}, altitude and hypoxia \cite{sim2019iron}, prolonged detraining conditions like injury-related immobility \cite{gledhill1985influence} and low iron intake and iron supplementation \cite{vsmid2024effects}. The change in iron can also be associated with pathological effects, due to liver disease, chronic inflammatory conditions with malabsorption (such as inflammatory bowel disease or celiac disease \cite{mahadea2021iron}), hereditary hemochromatosis, thyroid disorders, blood loss, or vitamin B12 deficiency. Medical treatments can lead to blood loss and iron-deficiency anemia \cite{tai2021non}. In particular, gastrointestinal bleeding is another potential reason for reductions of total hemoglobin mass (and thus ferritin) in athletes, especially in ultramarathon athletes which are reported to have a high prevalence for gastroinintestinal bleeding \cite{baska1990gastrointestinal}. This was also demonstrated with the follow-up of one elite rower in \cite{treff2014impact}. More generally, Nabhan et al. showed that iron deficiency is significantly lower in athletes compared to non-athletes \cite{nabhan2020serum}. Completing this observation, Lippi et al. showed that the serum ferritin concentrations of professional endurance athletes (cross-country skiers and road cyclists) are 2-fold to 3-fold higher than those of matched sedentary individuals and amateur athletes (road cyclists) \cite{lippi2005serum}. It is also possible that seasonality such as training and competitions periods vs. less intense periods affect ferritin concentrations. Similarly, previous investigations measured the effect of training and nutrition on the IGF1 biomarkers and showed that the training type (endurance vs. strength) \cite{khalid2020type}, the energy deficits (e.g. in ultra-endurance events) \cite{geesmann2017association}, and the protein and overall caloric intake \cite{gulick2020exercise} can affect the IGF1 dynamics with time, potentially explaining the observed proportions. It is also interesting to note that there are some similarities between the two datasets, as the high proportions of abnormality are detected in the same biomarkers. There also exist some differences between the two datasets, in particular the selected transformations may differ for the same biomarkers (see Tab. \ref{tab1:Normalite}). It underlines the importance of transforming the dataset based on the population of study and testing different transformations. This does not go against the individualized rationale of the proposed approach, as adjusting a different transformation per sequence (ie. each sequence would be tested for a transformation) would instead lead to overfitting.

The proposed approach can be refined for medical purposes. We detect all the abnormal sequences, including sequences generated from malfunctioning medical devices. For example we detect $4.45\%$ (soccer) and $4.75\%$ (cycling) of \textquoteleft constant\textquoteright sequences (ie. sequences containing the same value for all samples) in the hemoglobin samples among the sequences having at least 3 samples ($n \ge 3$). Other biomarkers also have constant sequences, with the maximum proportion of $1.81\%$ for the erythrocytes in soccer and $2.21\%$ for the reticulocytes (\%) in cycling. However, we do not have the information about the medical devices, nor about the athlete\textquoteright s training conditions, pathologies or medications, nutritional strategies, environmental factors, etc. They are known to affect biological markers values \cite{pedlar2019blood, sim2019iron}. It is possible to capture these effects using more sophisticated models, as we demonstrated using the model B and C if the data is available.

There are some limitations to the presented approach. First, there could be some intra-individual correlation between the variables $X_i$ as some collections were only separated by a few days \cite{buoro2018short, cooper1996day, sun2016reproducibility}. Of course, this depends on the biomarker and the intraclass correlation coefficient is limited or unavailable in the literature for several biomarkers (hemoglobin, hematocrit levels, reticulocytes, osteocalcin). This can limit the descriptive analysis presented here. Second, we did not have any information about input errors which can occur during the data collection. This could also affect the detection rates. However, the proposed framework (Fig. \ref{fig1:blockdiagram}) remains flexible enough to integrate new information and to further capture biological complexity using more sophisticated model designs. In line with this, the improvement of the proposed Z-score methods would allow to increase the statistical power of the detection while leading to a better understanding of the features affecting intra-individual variance in the personalized medicine context. It could also be used as a decision-assistance tool to automate the screening detection of athlete’s biological follows-up and to highlight possible pathological conditions. Applying this automatic screening to larger or specific cohorts could help understand the biological conditions of elite athlete’s regarding their specificities.

\section{Acknowledgments}
We thank Arthur Simon for providing additional help with the DevianLM package.

\bibliographystyle{unsrt}
\bibliography{bibfile_Zscore}

\end{document}